\documentclass[conference]{IEEEtran}
\IEEEoverridecommandlockouts
\usepackage{cite}
\usepackage{amsmath,amssymb,amsfonts}
\usepackage{algorithmic}
\usepackage{graphicx}
\usepackage{textcomp}
\usepackage{xcolor}
\usepackage{listings}
\usepackage{csquotes}
\usepackage{comment}
\usepackage{todonotes}
\usepackage{hyperref}
\usepackage{booktabs}
\usepackage{caption}

\usepackage[switch,columnwise]{lineno}

\definecolor{commentgreen}{RGB}{2,112,10}
\definecolor{eminence}{RGB}{108,48,130}
\definecolor{weborange}{RGB}{255,165,0}
\definecolor{frenchplum}{RGB}{129,20,83}

\lstdefinestyle{numbers} {numbers=none, stepnumber=1, numberstyle=\tttfamily, numbersep=5pt}
\lstdefinestyle{MyFrame}{backgroundcolor=\color{yellow!8},frame=shadowbox, breaklines=true}
\lstdefinestyle{myFontStyle}{commentstyle=\color{commentgreen}\tttfamily, morekeywords={CudaMalloc, CudaMallocManaged, __global__}}

\lstdefinestyle{cudacalls} {language=C, style=numbers, style=MyFrame, frame=lines, style=MyFontStyle}

\def\BibTeX{{\rm B\kern-.05em{\sc i\kern-.025em b}\kern-.08em
    T\kern-.1667em\lower.7ex\hbox{E}\kern-.125emX}}

\begin{document}

\title{Rapid Exploration of Optimization Strategies on Advanced Architectures using TestSNAP and LAMMPS \\
\thanks{Sandia National Laboratories is a multimission laboratory managed and operated by National Technology and Engineering Solutions of Sandia, LLC., a wholly owned subsidiary of Honeywell International, Inc., for the U.S. Department of Energy’s National Nuclear Security Administration under contract DE-NA-0003525.
This paper describes objective technical results and analysis. Any subjective views or opinions that might be expressed in the paper do not necessarily represent the views of the U.S. Department of Energy or the United States Government.
}
}

\author{\IEEEauthorblockN{1\textsuperscript{st} Rahulkumar Gayatri}
\IEEEauthorblockA{\textit{NERSC} \\
\textit{Lawrence Berkeley National Lab}\\
Berkeley, USA \\
rgayatri@lbl.gov}
\and
\IEEEauthorblockN{2\textsuperscript{nd} Stan Moore}
\IEEEauthorblockA{\textit{Sandia National Laboratories} \\
Albuquerque, USA \\
stamoor@sandia.gov}
\and
\IEEEauthorblockN{3\textsuperscript{rd} Evan Weinberg}
\IEEEauthorblockA{\textit{NVIDIA Corporation}\\
Santa Clara, USA \\
eweinberg@nvidia.com}
\and
\IEEEauthorblockN{4\textsuperscript{th} Nicholas Lubbers}
\IEEEauthorblockA{\textit{Los Alamos National Lab} \\
Los Alamos, USA \\
nlubbers@lanl.gov}
\and
\IEEEauthorblockN{5\textsuperscript{th} Sarah Anderson}
\IEEEauthorblockA{\textit{Cray Inc} \\
Minneapolis, USA \\
saraha@hpe.com}
\and
\IEEEauthorblockN{6\textsuperscript{th} Jack Deslippe}
\IEEEauthorblockA{\textit{NERSC} \\
\textit{Lawrence Berkeley National Lab}\\
Berkeley, USA \\
jrdeslippe@lbl.gov}
\and
\IEEEauthorblockN{7\textsuperscript{th} Danny Perez}
\IEEEauthorblockA{\textit{Los Alamos National Lab} \\
Los Alamos, USA \\
danny\_perez@lanl.gov}

\and
\IEEEauthorblockN{8\textsuperscript{th} Aidan P. Thompson}
\IEEEauthorblockA{\textit{Sandia National Laboratories} \\
Albuquerque, USA \\
athomps@sandia.gov}
}

\maketitle

\begin{abstract}
The exascale race is at an end with the announcement of the Aurora and Frontier machines.
This next generation of supercomputers utilize diverse hardware architectures to achieve their compute performance, providing an added onus on the performance portability of applications.
A solution to this challenge is the evolution of performance-portable frameworks, providing unified models for mapping abstract hierarchies of parallelism to diverse architectures.
Kokkos is one such performance portable programming model for C++ applications, providing back-end implementations for each major HPC platform.
Even with the availability of performance portable frameworks, modern compute resources which feature heterogeneous node architectures containing multicore processors and manycore accelerators, challenge application developers to restructure algorithms to expose higher degrees of parallelism.

The Spectral Neighbor Analysis Potential (SNAP) is a machine-learned inter-atomic potential utilized in cutting-edge molecular dynamics simulations.
Previous implementations of the SNAP calculation showed a downward trend in their performance relative to peak on newer-generation CPUs and low performance on GPUs.
In this paper we describe the restructuring and optimization of SNAP as implemented in the Kokkos CUDA backend of the LAMMPS molecular dynamics package, benchmarked on NVIDIA GPUs.
We identify novel patterns of hierarchical parallelism, facilitating a minimization of memory access overheads and pushing the implementation into a compute-saturated regime.
Our implementation via Kokkos enables recompile-and-run efficiency on upcoming architectures.
We find a $\sim$22x time-to-solution improvement relative to an existing implementation as measured on an NVIDIA Tesla V100-16GB for an important benchmark.
\end{abstract}


\section{Introduction} \label{sec:intro}
In order to reach \textit{exaflop} performance while maintaining power constraints, heterogeneous node architectures have become the norm.
The heterogeneous node combination consists of one or more \textit{host} CPUs, which are usually multicore processors, and one or more accelerators or GPUs also called a \textit{device}.
As of June 2020, six out of the top ten machines in the list of top 500 supercomputers depend on such heterogeneous architectures to achieve their compute performance \cite{top500}.
The next generation of supercomputers such as Perlmutter, Aurora, and Frontier follow the same path by relying on NVIDIA, Intel, and AMD GPUs, respectively, for the majority of their compute bandwidth.

The trend of heterogeneous node composition has also led to an increased focus on application development.
Application developers are now forced to re-think their parallelization strategies to effectively utilize nearly a 100$\times$ increase in the number of threads available on each node.
Apart from effectively utilizing the increase in the computational power, programmers now need to parallelize the application code for both multicore and manycore architectures.
This has led to the development of multiple performance-portability frameworks such as Kokkos \cite{kokkos}, Raja \cite{raja}, OpenMP offload, etc.
The intention of these models is to provide a single coding front-end for the application developers while the frameworks maintain multiple backend implementations optimized for specific hardware architectures such as CPUs and GPUs.

While these frameworks strive to generate optimized code tailored for the underlying architecture, it is also important for the application developers to restructure their code to exploit the massive increase in computational power via hardware tailored to exploiting parallelism.
The threads on a GPU are organized in a hierarchical fashion. For example on NVIDIA GPUs, $\mathcal{O}(100-1000)$ threads are grouped as a thread block and share resources among themselves.
Similarly on Intel's Gen9 GPU architectures, the threads are divided into threadgroups and threads.
Application developers need to design a parallel implementation that can distribute work effectively across both levels of the hierarchy.

Additionally, GPUs have their own memory space and memory hierarchies, increasing the complexity of the application.
On GPUs, threads performing fully coalesced memory reads (i.e., access of consecutive memory locations) minimize memory transactions and thus minimize latencies.
Such a memory access pattern should be avoided on CPUs since separate threads accessing memory locations in the same cache line results in false sharing and thrashing.

%
%
%
Compared to CPUs, modern GPUs have a very large ratio of compute throughput to memory bandwidth, also referred to as a high arithmetic intensity (AI).
Optimally written kernels on CPUs often become memory bound on GPUs. 
In these scenarios new strategies which reduce reads are invaluable: kernel fusion and even redundant computation can be net-beneficial if such optimizations improve the AI of a kernel~\cite{kernel-fusion}.

Spectral Neighbor Analysis Potential (SNAP) is a computationally intensive interatomic potential in the LAMMPS \cite{Plimpton1995} molecular dynamics software package.
With the introduction of new CPU and GPU architectures, SNAP's performance relative to the peak flop rating of the architecture showed a downward trend on CPUs and low compute utilization on GPUs.
In this paper we describe our efforts to improve the performance of SNAP by optimizing on all the points mentioned above.
We explain how we overcame performance deterrents in SNAP to gain a $\sim$22$\times$ performance increase over the existing GPU implementation.

\section{The SNAP Force Kernel} \label{sec:SNAP}

\newcommand{\bu}{{\bf u}}
\newcommand{\bdu}{{\bf du}}
\newcommand{\bU}{{\bf U}}
\newcommand{\bZ}{{\bf Z}}
\newcommand{\bY}{{\bf Y}}

Interatomic potentials (IAPs) are a critical part of any classical molecular dynamics (MD) simulation.
The use of a classical IAP implies that accuracy-limiting approximations are acceptable.
The most important assumptions shared by many IAPs are as follows.
First, all-electron mediated interactions of atoms can be described by the Born-Oppenheimer potential energy surface.
Second, the force on a given atom does not depend on the positions of atoms beyond a certain distance.
This latter constraint permits the use of efficient algorithms that are linear scaling in the number of atoms, as well as ensuring that large
problems can be efficiently distributed over leadership computing platforms.
Many IAPs capture local interactions using approximations inspired by known physical and chemical phenomena, such as chemical bonding, electrostatic screening, local coordination, etc.
The development of new IAPs follows a decades long trend where much of the effort has focused on more accurate, but also more complex and computationally expensive IAPs {\cite{Plimpton2012}.

A recent branch of this development combines strategies of MD and data science, producing machine-learned (ML) IAPs.
Machine-learned IAPs translate the atomic neighborhood into a set of generalized \textit{descriptors}.
These descriptors are independently weighted to match a database of higher fidelity results (e.g. \emph{ab initio} quantum electronic structure calculations).
A variety of different descriptors which describe the local environment of an atom exist in the literature~\cite{Bartok2013}, such as symmetry functions~\cite{Behler2007}, bispectrum components\cite{Bartok2010}, and the Coulomb matrix\cite{Rupp2012}.
The Spectral Neighbor Analysis Potential (SNAP)~\cite{Thompson2015} utilizes a basis expansion of bispectrum components as a descriptor of atomic environments.
Drautz has recently shown that many of these descriptors, including the SNAP bispectrum, share a common mathematical foundation in the atomic cluster expansion for the Born-Oppenheimer potential energy function\cite{Drautz2019}.

\subsection{Mathematical Structure of SNAP}

In the SNAP potential energy model, the total energy of a configuration of atoms is composed as a sum of atomic energies.  For each atom $i$, its energy $E_i$ is assumed to be a function of the positions of neighbor atoms out to some finite distance $R_{\mathrm{cut}}$.  The positions of neighbor atoms are represented as an atomic density function defined in the 3D ball of radius $R_{\mathrm{cut}}$. In SNAP, this compact domain is mapped to the unit 3-sphere.  Expanding the neighbor density as a Fourier series,  we obtain the following Fourier coefficients
\begin{eqnarray}
\label{eqn:u}
\bU_j &=& \sum_{r_{ik} < R_{\mathrm{cut}}} {f_c(r_{ik}) \bu_j(\theta_0,\theta,\phi)}
\end{eqnarray}
where $\bU_j$ are the Fourier expansion coefficients and $\bu_j$ are hyperspherical harmonics on the unit 3-sphere.  Both $\bU_j$ and $\bu_j$ are rank $(2j+1)$ complex square matrices, where the index $j$ takes half-integer values $\{0, \frac{1}{2},1,\frac{3}{2},\ldots\}$.  The 3D vector ${\bf r}_{ik} = {\bf r}_k - {\bf r}_i$ is the position of neighbor atom $k$ relative to the central atom $i$ and is mapped to a point on the 3-sphere given by the three polar coordinates $\theta_0$, $\theta$, and $\phi$.  The sum is over all neighbor atoms within the cutoff.  The switching function $f_c(r)$ ensures that contributions go smoothly to zero as $r$ approaches $R_{\mathrm{cut}}$.   For conciseness of presentation we omit density weighting factors and self-contributions which require negligible computation but are important for constructing physically realistic potentials. Full details are given in~\cite{Thompson2015}.

The matrices $\bU_j$ are complex-valued and are not directly useful as descriptors because they are not invariant under rotation of the polar coordinate frame.  However, the following scalar triple products of matrices are real-valued and invariant under rotation~\cite{Bartok2010}:
\begin{eqnarray}
\label{eqn:z}
B_{j_1j_2j}  &=& \bU_{j_1} \otimes_{j_1j_2}^j \bU_{j_2} \colon \bU_j^* \\
&=& \bZ_{j_1j_2}^j  \colon \bU_j^*
\end{eqnarray}

The symbol $\otimes_{j_1j_2}^j$ indicates a Clebsch-Gordan product of matrices of rank $2j_1+1$ and $2j_2+1$ yielding a matrix of rank $2j+1$, which we define here to be $\bZ_{j_1j_2}^j$.  The computational complexity of this product is $\mathcal{O}(j^4)$.  The $\colon$ symbol indicates the element-wise scalar product of two matrices of equal rank, an operation of computational complexity $\mathcal{O}(j^2)$.   The resultant real scalar bispectrum components $B_{j_1j_2j}$ characterize the strength of density correlations at three points on the 3-sphere.  The lowest-order components describe the coarsest features of the density function, while higher-order components reflect finer detail.  The bispectrum components defined here have been shown to be closely related to the 4-body basis functions of the Atomic Cluster Expansion introduced by Drautz~\cite{Drautz2019}.  In SNAP, we assume that the local energy can be expressed as a linear function of all the distinct bispectrum components formed from matrices $\bU_j$ up to some maximum degree $J$.  We enumerate the bispectrum components by restricting $0 \le 2j_2 \le 2j_1 \le 2j \le 2J$, so that the number of unique bispectrum components scales as $\mathcal{O}(J^3)$. The factor of 2 is a convenient convention to avoid half-integers.

For a particular choice of $J$, we can list the $N_B$ total bispectrum components in some arbitrary order as ${B}_{1},\ldots,{B}_{N_B}$, atom index $i$ implicit, and express the energy as a linear function of these
\begin{equation}
\label{eqn:snapatomenergy}
E_i({\bf B})  = \sum_{l=1}^{N_B} \beta_l B_l
\end{equation}
where $ \beta_l$ are the linear SNAP coefficients. These coefficients are trained via ML methods to define the SNAP energy model for a particular material.
The force on each atom $k$  is obtained by summing over all neighbor atoms $i$ and all bispectrum components
\begin{equation}
\label{eqn:snapforce1}
{\bf F}_k  = -  \sum_{i=1}^N \sum_{l=1}^{N_B} \beta_l \frac{\partial B_l}{\partial {\bf r}_k}
\end{equation}

As described in Ref.~\citenum{Thompson2015}, the partial derivative of $B_l$ w.r.t. ${\bf r}_k$ is a sum of three terms, each involving a neighbor-atom independent and thus precomputable $\bZ$ and a neighbor-atom dependent derivative of $\bU$,
\begin{eqnarray}
\label{eqn:dbidrk}
\frac{\partial B_{j_1j_2j}}{\partial {\bf r}_k} &=& \bZ_{j_1j_2}^j \colon \frac{\partial \bU_j^*}{\partial {\bf r}_k}
\\ \nonumber&&
+ \bZ_{jj_2}^{j_1} \colon \frac{\partial \bU_{j_1}^*}{\partial {\bf r}_k} + \bZ_{jj_1}^{j_2} \colon \frac{\partial \bU_{j_2}^*}{\partial {\bf r}_k}
\end{eqnarray}

This formulation prescribes the structure of an algorithm for computing forces from the SNAP potential. We present this in Listing~\ref{lst:SNAP-code}.

\subsection{Initial SNAP Pseudocode}
Listing \ref{lst:SNAP-code} shows the pseudocode of SNAP implementation by correlating the routines with the equations shown above.
Initially the \emph{build\_neighborlist} routine generates a list of neighbor atoms within the cutoff distance $R_{\mathrm{cut}}$.
Next, in the \emph{compute\_U} routine, we calculate the expansion coefficients $\bU_j$ from (\ref{eqn:u}) for each atom and neighbor pair and store them in \emph{Ulist}. The sum of these coefficients over neighbors is stored in \emph{Ulisttot}.
Given \emph{Ulisttot}, in the \emph{compute\_Z} routine we calculate the Clebsch-Gordan product based on (\ref{eqn:z}) and store the results of $\bZ_{j_1j_2}^j$ in the \emph{Zlist} structure, a 5-dimensional array of complex double precision values.
The computational complexity per atom of \emph{compute\_U} and \emph{compute\_Z} are $\mathcal{O}(J^3 N_{\mathrm{nbor}})$ and $\mathcal{O}(J^7)$, respectively.
We next enter a nested loop over neighbors.
In the \emph{compute\_dU} routine we compute $dU$, the derivative of $\bU$ w.r.t. the position of one neighbor atom ${\bf r}_k$, storing the results in the \emph{dUlist} data structure.
Subsequently in the routine \emph{compute\_dB} we compute the partial derivatives of $B_l$ shown in (\ref{eqn:dbidrk}), storing the results in~\emph{dBlist}.
The computational complexity per neighbor atom of \emph{compute\_dU} and \emph{compute\_dB} are $\mathcal{O}(J^3)$ and $\mathcal{O}(J^5)$, respectively.
Last, in the \emph{update\_forces} routine, we compute the force contribution due to the neighbor as shown in (\ref{eqn:snapforce1}).
This final operation has a computational complexity per atom-neighbor pair of $\mathcal{O}(J^3)$.
We emphasize that as formulated, the order of the functions cannot be changed because the outputs from one routine pipe into the following routine.
\begin{minipage}{\linewidth}
\begin{lstlisting} [caption={SNAP code}, captionpos=b, label=lst:SNAP-code]
for(int natom=0; natom<num_atoms; ++natom)
{
  // build neighbor-list for each atom
  build_neighborlist();
  // compute atom specific coefficients
  compute_U(); //Ulist and Ulisttot
  compute_Z(); //Zlist
  // For each (atom,neighbor) pair
  for(int nbor=0; nbor<num_nbors; ++nbor)
  {
    compute_dU(); //dUlist
    compute_dB(); //dBlist
    update_forces(); //force-array
  }
}
\end{lstlisting}
\end{minipage}

\subsection{GPU implementation of SNAP}
The original GPU implementation of SNAP used the Kokkos~\cite{kokkos} framework to distribute the work described in Listing~\ref{lst:SNAP-code} across the threads of a GPU.  This implementation was based on prior work by Moore and Trott\cite{Moore2018}, which in turn was based on an earlier CUDA implementation of SNAP\cite{Trott2014}.
The loop over atoms on line 2 of listing~\ref{lst:SNAP-code} is mapped to a loop over Kokkos \emph{teams}, which are an abstraction of CUDA thread blocks.
Further parallelism over neighbors, line 12 of listing~\ref{lst:SNAP-code}, and over bispectrum components, implicit in \emph{compute\_[U,Z,dU,dB]}, is mapped to Kokkos's hierarchical parallelism. This includes the \emph{TeamThread} abstraction, parallelism over Kokkos ``threads'', and the \emph{ThreadVector} abstraction, parallelism over ``vector lanes''. In the case of GPUs these abstractions can map to warps and threads within warps (a ``vector width'' of 32), respectively.



Table~\ref{tab:snap1} lists the performance of these initial implementations of the SNAP potential across several HPC architectures. 
The problem sizes chosen for these comparisons comprised of 2000 atoms with 26 neighbors per atom and $2J=8$.
The Kokkos version of SNAP was used for the GPU benchmarks; the original (non-threaded) SNAP version was used for all others.

\begin{table}
  \caption{SNAP performance on different hardware.}
  \label{tab:snap1}
  \begin{center}
  \resizebox{\columnwidth}{!}{
    \begin{tabular}{|ll||ccc|}
\toprule
& & Speed & Peak/node & Fraction of Peak \\
Hardware & Year & (Katom-steps/s) & (Tflops) & (normalized) \\
\midrule
Intel SandyBridge & 2012 & 17.7 & 0.332 & 1.0  \\
IBM PowerPC       & 2012 & 2.52 & 0.205 & 0.23   \\
AMD CPU           & 2013 & 5.35 & 0.141 & 0.71  \\
NVIDIA K20X       & 2013 & 2.60 & 1.31  & 0.037 \\
Intel Haswell     & 2016 & 29.4 & 1.18  & 0.47  \\
Intel KNL         & 2016 & 11.1 & 2.61  & 0.080 \\
NVIDIA P100       & 2016 & 21.8 & 5.30  & 0.077 \\
Intel Broadwell   & 2017 & 25.4 & 1.21  & 0.39  \\
NVIDIA V100   & 2018 & 32.8 & 7.8  & 0.079  \\
\bottomrule
    \end{tabular} }
  \end{center}
\end{table}

Performance speed for classical MD simulations is often reported in units of Katom-timesteps per second.
For example, given a 2000 atom system, the speed of 29.4~Katom-steps/s on Intel Haswell implies that the simulation rate was $\sim$15 MD timesteps per second.
The \emph{peak/node} column is the nominal maximum FLOP rate (double precision) for one CPU node or one GPU.
The \emph{fraction of peak} column is ratio of \emph{speed} divided by \emph{peak/node} for a particular platform \emph{relative to the Intel SandyBridge baseline}.
This convention is chosen to abstract away technical differences in FLOP counts between architectures, for example due to recomputing values on a GPU as opposed to loading from memory on a CPU.

The results in Table~\ref{tab:snap1} support our assertion that relative performance has declined with advances in CPU architectures.
Performance is low on GPUs despite the utilization of hierarchical parallelism and scratch memory in Kokkos.
To address these performance issues and arrive at a new parallelization strategy we created a proxy application \enquote{TestSNAP}, a stand-alone serial application reproducing the implementation given in listing~\ref{lst:SNAP-code}, without the additional complexities of a full molecular dynamics code (\url{https://github.com/FitSNAP/TestSNAP}).
%

In this paper we will systematically describe our parallelization and optimization process, benchmarking our progress in TestSNAP and LAMMPS relative to the initial Kokkos SNAP implementation, or \enquote{baseline}  \cite{LAMMPS_baseline} in LAMMPS.
The results shown in this paper are for systems of 2000 atoms with 26 neighbors each.
We consider two values of $J$, 8 and 14, corresponding to 55 and 204 bispectrum components, respectively.
We will henceforth use $2J8$ and $2J14$ as shorthand for these problem sizes.
The optimizations here are targeted towards NVIDIA's V100 GPU, although in most cases they are generic optimizations that are applicable to all GPUs.
All performance measurements given in this paper are from one NVIDIA V100 GPU of the Summit supercomputer at Oak Ridge National Laboratory.
\section{TestSNAP } \label{sec:TestSNAP}
The intention behind TestSNAP is to provide a testbed in which many different optimizations can be explored without needing to build and run the full LAMMPS code, allowing developers to
focus on the core components of SNAP algorithm and their implementation.
Successful optimization strategies can then be merged back into the LAMMPS production code.
There is no fundamental reason why these performance explorations could not have been performed directly in LAMMPS, however TestSNAP allowed us to effectively collaborate between our diverse team of developers irrespective of their familiarity to LAMMPS.
\subsection{Refactor compute routines}
One of the main disadvantages of the baseline implementation was the over-subscription of limited resources such as registers, leading to a limit on available occupancy.
This was a side effect of the use of a single large kernel being launched.
To address this, our first step was to refactor the SNAP algorithm into individual stages, as demonstrated in listing~\ref{lst:refactor-TestSNAP-code}.
%
\begin{minipage}{\linewidth}
\begin{lstlisting} [caption={Refactored TestSNAP code}, captionpos=b, label=lst:refactor-TestSNAP-code]
// build neighbor-list for all atoms
for(int natom=0; natom<num_atoms; ++natom)
  build_neighborlist();

  // compute matrices for all atoms
for(int natom=0; natom<num_atoms; ++natom)
  compute_U(); //Ulist(num_atoms,...)

for(int natom=0; natom<num_atoms; ++natom)
  compute_Z(); //Zlist(num_atoms,...)

  // For each (atom,neighbor) pair
for(int natom=0; natom<num_atoms; ++natom)
  for(int nbor=0; nbor<num_nbors; ++nbor)
    compute_dU(); //dUlist(num_atoms,...)

for(int natom=0; natom<num_atoms; ++natom)
  for(int nbor=0; nbor<num_nbors; ++nbor)
    compute_dB(); //dBlist(num_atoms,...)

for(int natom=0; natom<num_atoms; ++natom)
  for(int nbor=0; nbor<num_nbors; ++nbor)
    update_forces();
\end{lstlisting}
\end{minipage}
In this refactoring each stage can be viewed as a single GPU kernel which performs the necessary work for all atoms.
Launching individual kernels allows parameters such as block size to be specifically tailored to each kernel.
For some kernels register usage will be lower, leading to increased occupancy.
The order of loops for individual kernels can also be optimized.

This refactoring into separate kernels does have disadvantages.
The dominant disadvantage is that, because the state of memory does not persist across kernel calls, we need to manually ``cache'' results between kernel launches.
Every data structure now has an additional dimension to reference individual atoms as shown in comments of listing \ref{lst:refactor-TestSNAP-code}.
This increases memory requirements by a factor of the number of atoms being processed, in our case 2000.
While not a prohibitive issue, this did lead to novel challenges whose solutions will be discussed later in the paper.

We used the Kokkos framework to port TestSNAP to GPU in order to be consistent with the original baseline implementation.
For the rest of the paper we show code snippets from the Kokkos implementation of TestSNAP to explain our optimization strategies.
As a step towards transitioning to a Kokkos implementation, we pushed the atom and, as appropriate, neighbor loops inside the individual routines.
In addition, we converted the data structures \emph{Ulist,Zlist,dUlist,dBlist} to Kokkos \emph{views}, abstractions of multi-dimensional arrays.
%

\subsection{GPU implementation}

In this subsection we describe the early GPU implementations of the refactored TestSNAP code. 
\subsubsection{Atom loop parallelization}
\label{sec:atom-loop-par}
As with the baseline implementation, the first step is to parallelize over independent atoms.
In our new implementation, we assign one GPU thread to each atom, in contrast to the baseline where an entire thread block was assigned to each atom.
We utilize this pattern for each of the four dominant kernels, \emph{compute\_[U,Z,dU,dB]}.
As part of our refactoring with Kokkos each complex data structure in listing~\ref{lst:refactor-TestSNAP-code} is converted into Kokkos views of type \emph{SNAcomplex}, a thin wrapper of a complex double.
Listing \ref{lst:compute_u} sketches the refactoring of \emph{compute\_U} to use Kokkos, as well as the definition of \emph{SNAcomplex}.
The salient feature of this refactoring is that the per-atom loop body within \emph{compute\_U} has been wrapped inside a C++ lambda.
This lambda is passed to the \emph{Kokkos::parallel\_for} construct along with the loop dimension.
\begin{minipage}{\linewidth}
\begin{lstlisting} [caption={\emph{Kokkos::parallel\_for} routine to distribute work of atoms inside compute\_U routine}, captionpos=b, label=lst:compute_u]
void compute_U
{
  struct SNAcomplex {double re, im;};
  using Kokkos::parallel_for;
  parallel_for(num_atoms, KOKKOS_LAMBDA(const int natom)
  {
    Ulist(natom,...) = ...
  });
}
\end{lstlisting}
\end{minipage}

Broadly, Kokkos then manages dispatching work to the target architecture.
Here, Kokkos distributes the per-atom work across GPU threads, only offering the guarantee that the lambda body is executed once per atom on the GPU.
The SNAP algorithm is well suited to this design because each atom performs independent work.
We will present the results of distributing the work of an atom across threads of a GPU along with the next optimization since it is a logical extension to our initial step.
%

\subsubsection{Atom and neighbor loop parallelization}
\label{sec:neigh-loop-par}
For our next optimization we exposed the additional parallelism across neighbors inside the \emph{compute\_dU} and \emph{compute\_dB} routines.
Kokkos provides an \emph{MDRangePolicy} as an abstraction for parallelizing over nested loops, the use of which we sketch in listing~\ref{lst:compute-du-collapse}.
%
%
\begin{minipage}{\linewidth}
\begin{lstlisting} [caption={\emph{MDRangePolicy} to collapse atom and neighbor loops in compute\_dU routine.}, captionpos=b, label=lst:compute-du-collapse]
void compute_dU
{
  using Execspace = Kokkos::execution_space;
  using MDPolicyType2D = typename Kokkos::MDRangePolicy<ExecSpace, Kokkos::Rank<2>, int>;
  MDPolicyType2D mdPolicy2D({0, 0}, {num_atoms, num_nbor});
  parallel_for(mdPolicy2D, KOKKOS_LAMBDA(const int natom, const int nbor)
  {
    dulist(natom,nbor) = ...;
  });
}
\end{lstlisting}
\end{minipage}

Line 3 of listing~\ref{lst:compute-du-collapse} acquires the default execution space for launching the work inside a Kokkos \emph{parallel\_for} construct, the NVIDIA GPU in the context of this work.
Line 4 shows the declaration of a 2D range policy type in the default execution space with a 32-bit integer iterator.
Line 5 creates a 2D policy type where the first and second loop are over \emph{num\_atoms} and \emph{num\_nbor} respectively.
Line 7 shows the launching of the 2D range policy.

\begin{figure}
\centering
\includegraphics[width=1.0\columnwidth]{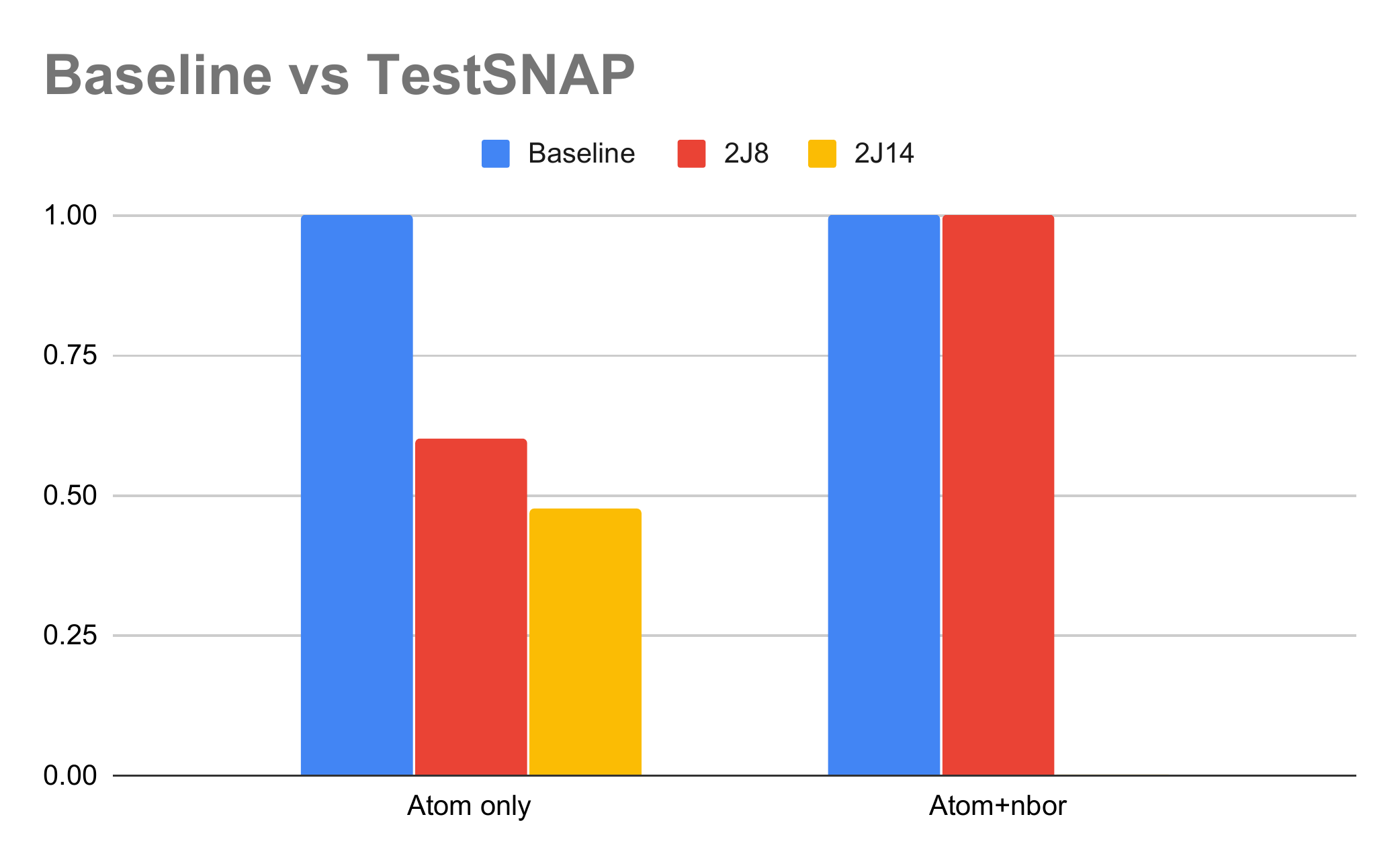}
\caption{Baseline performance compared with TestSNAP after atom and neighbor loop parallelization on V100.}
\label{fig:atom-nbor-loop}
\end{figure}

The results of the two parallelization strategies discussed above are shown in \figurename~\ref{fig:atom-nbor-loop} for the 2J8 and 2J14 problem sizes on NVIDIAs V100 architecture.
As explained earlier, we present the results of our GPU implementation relative to the baseline implementation.
Hence anything higher than \enquote{1} would imply performance improvement and conversely anything lower would mean that our implementation is slower than the baseline.

In \figurename~\ref{fig:atom-nbor-loop}, the first set of bars show the performance comparison of our implementation when only atom loop is distributed across threads of a GPU.
For our target problem size we see a 1.5$\times$ and 2.0$\times$ regression for the 2J8 and 2J14 problem sizes.
A regression at this stage is reasonable.
We increased our memory footprints to 3 and 5GB for the two problem sizes, respectively, with implications for cache reuse.
In addition, relative to the baseline we are exposing less parallelism by not threading over neighbors and $j$, $j_1$, $j_2$.
The latter is our next point of focus.

The second set of bars in \figurename~\ref{fig:atom-nbor-loop} show the performance comparison of atom and neighbor loop parallelization.
Refactoring to use an \emph{MDRangePolicy} increases the amount of exposed parallelism by a factor equal to the number of neighbors per atom, here 26.
Unfortunately, we now also need to store information between kernel launches as a function of both atom and neighbor number.
This increases the memory footprint of \emph{[U,dU,dE]list} by a number-of-neighbors factor, again 26, leading to a total memory footprint of 5 GB for the 2J8 problem size and an out-of-memory error for the 2J14 problem size!
Hence as can be observed in the second set of bars in the figure, while the performance parity for 2J8 is restored with the baseline code, the comparison for 2J14 is conspicuously missing. For comparison, the baseline code has a GPU memory footprint of 2 GB for the 2J8 problem and 14 GB for the 2J14 problem.

%
%

There is no trivial solution to the out-of-memory error for the 2J14 problem size.
The robust solution to this problem is given by the so-called adjoint refactorization which we describe in section~\ref{sec:y-array}.
%

\section{Adjoint Refactorization} \label{sec:y-array}
The original formulation of the SNAP force calculation relied on pre-calculating and storing the $\bZ$ matrices for each atom.
This avoided repeated calculation of the $\mathcal{O}((2j+1)^4)$ Clebsch-Gordan products for each of the $(2J+1)$~$\bZ$ matrices.
With this strategy the total memory footprint per atom scales as $\mathcal{O}(J^5)$.
To avoid this issue we combine (\ref{eqn:snapforce1}) and (\ref{eqn:dbidrk}) and define a new quantity $\bY$ that is the adjoint of $\bf B$ with respect to $\bU$,
\begin{eqnarray}
\label{eqn:y}
\bY_j   &=&  \sum_{j_1j_2} \beta_{j_1j_2}^j  \bZ_{j_1j_2}^j\;.
\end{eqnarray}

In this formulation each $\bZ$ matrix can be computed and immediately accumulated to the corresponding $\bY_j$.
This reduces the $\mathcal{O}(J^5)$ storage requirement for $\bZ$, replacing it by the $\mathcal{O}(J^3)$ storage requirement for $\bY_j$.
As noted in a recent paper by Bachmayr \emph{et al.}~\cite{bachmayr2019approximation}, this refactorization is equivalent to the backward differentiation method for obtaining gradients from neural networks.
This separate computation of $\bY_j$ has the additional benefit of eliminating the sum over $j_1$ and $j_2$ from \emph{compute\_dB}.
Since $\bY_j$ is neighbor-independent this eliminates an additional $\mathcal{O}(N_{nbor})$ of storage and computation relative to the previous implementation.

With this refactorization we can avoid calculating and storing \emph{dB} prior to the force calculation, an $\mathcal{O}(J^3)$ reduction in memory overheads.
The optimized SNAP force calculation is now formulated as a sum over one bispectrum index $j$ instead of three, giving
\begin{equation}
\label{eqn:snapforce2}
{\bf F}_k  = - \sum_{i=1}^N \sum_{j=0}^J \bY_j \colon \frac{\partial \bU_j^*}{\partial {\bf r}_k}.
\end{equation}

In practice we store the force contributions to an $N_{atom} \times N_{neigh}$ structure \emph{dElist}.
This is because in the full LAMMPS MD workflow individual \emph{dElist} components contribute to other quantities of interest, such as the virial tensor.

Listing \ref{lst:Yi-TestSNAP-code} shows the modified TestSNAP algorithm with the adjoint factorization implemented.
In summary, we have replaced the routines \emph{compute\_Z} and \emph{compute\_dB} by \emph{compute\_Y} and \emph{compute\_dE}, respectively.
\begin{minipage}{\linewidth}
\begin{lstlisting} [caption={TestSNAP code}, captionpos=b, label=lst:Yi-TestSNAP-code]
int natom, nbor;
build_neighborlist();
compute_U();
compute_Y();
compute_dU();
compute_dE();
update_forces();
\end{lstlisting}
\end{minipage}
%

\section{Optimization of Refactored Code} \label{sec:TestSNAP2}

The adjoint refactorization reduced both memory overheads and the computational complexity of the SNAP calculation.
We additionally flattened jagged multi-dimensional arrays related to the $j_1$, $j_2$, $j$ structures which further reduced memory use.
These optimizations reduced the memory requirements for the $2J=14$ problem size to 12 GB, rendering our algorithm tractable on a V100-16GB.
The adjoint refactorization and memory reduction improved performance on CPUs as well giving us a 3$\times$ performance boost on the Intel Broadwell CPU for the 2J8 problem size.


We document next a series of optimizations we performed on the refactored algorithm.
A summary of our figure of merit, the grind-time, relative to the baseline is given in \figurename~\ref{fig:TestSNAP-progress-2J8} and~\ref{fig:TestSNAP-progress-2J14} for the 2J8 and 2J14 problem sizes, respectively.
The performance numbers shown in the figures are obtained by running TestSNAP on NVIDIA's V100 GPU.
The labels on the x-axis correspond to subsection numbers, \emph{V1} through \emph{V7}, in which we provide detailed descriptions of our optimizations.
The height of the bar for any given subsection assumes the optimizations from all previous subsections are in place.
\begin{figure}
\centering
\includegraphics[width=1.0\columnwidth]{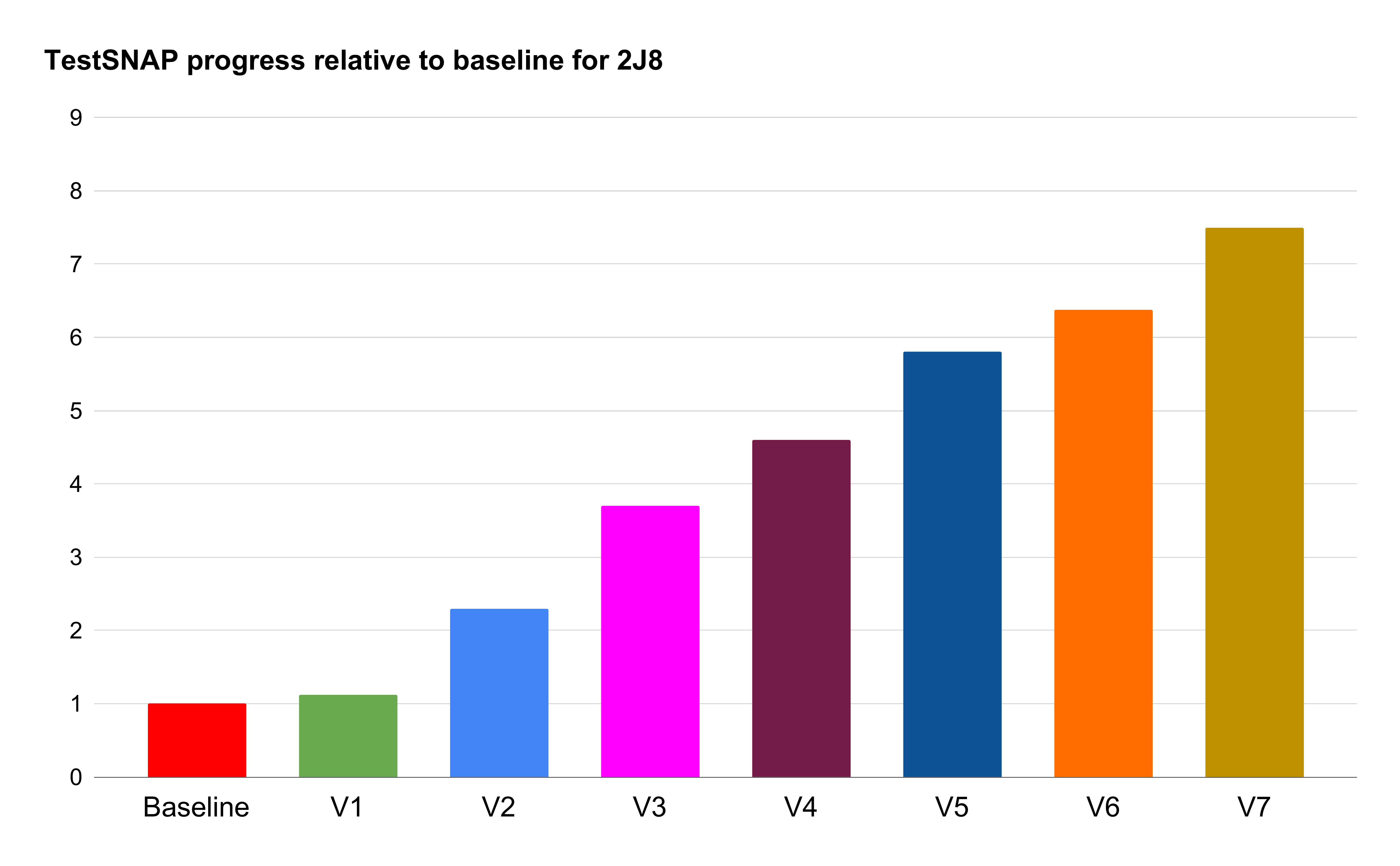}
\caption{TestSNAP progress relative to baseline for 2J8 problem size on NVIDIA V100.}
\label{fig:TestSNAP-progress-2J8}
\end{figure}
\begin{figure}
\centering
\includegraphics[width=1.0\columnwidth]{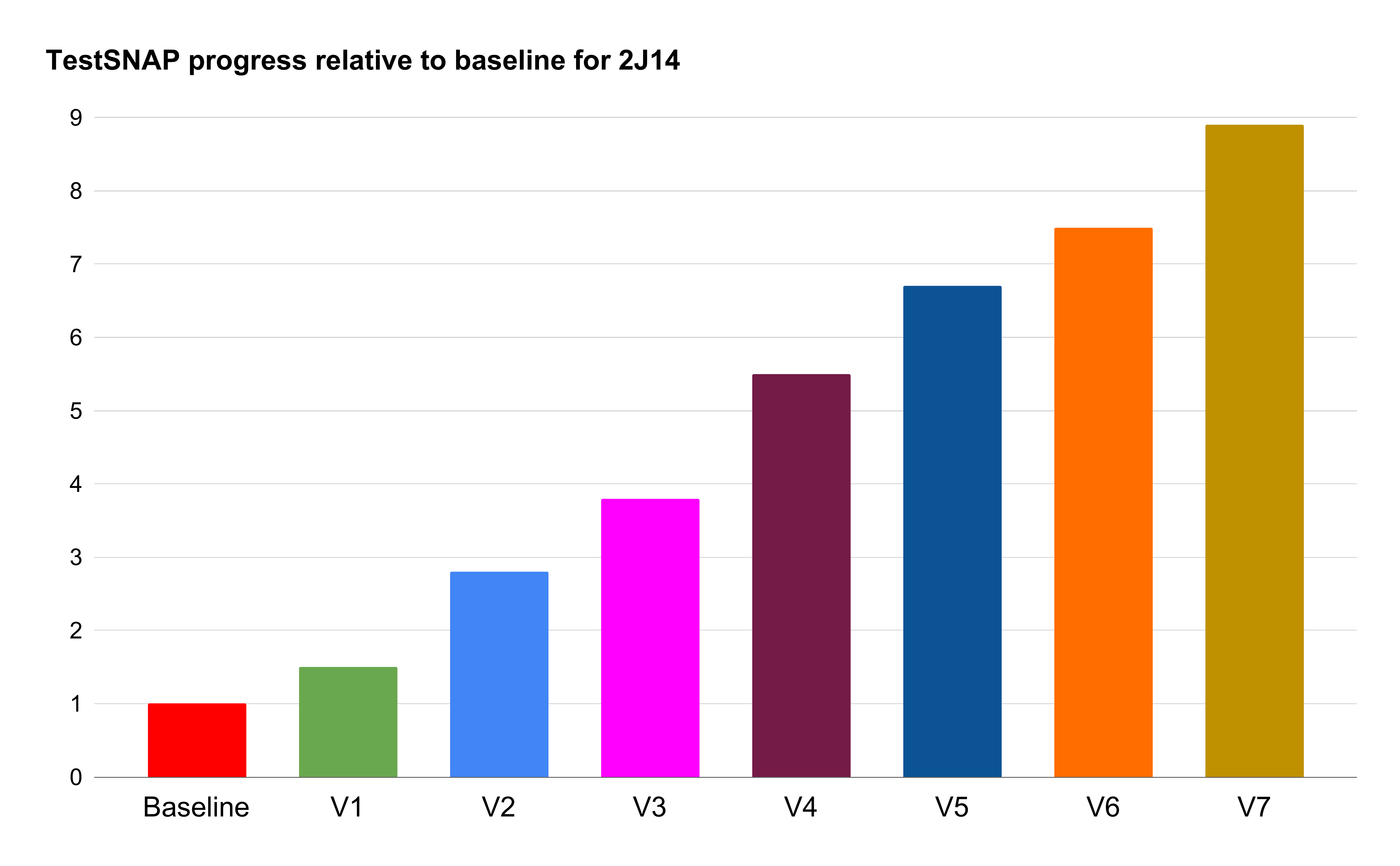}
\caption{TestSNAP progress relative to baseline for 2J14 problem size on NVIDIA V100.}
\label{fig:TestSNAP-progress-2J14}
\end{figure}

\subsection {\textit{V1} - Atom loop parallelization}
Following the pattern of our initial TestSNAP Kokkos implementation, our first step is to refactor the algorithm into four discrete kernels, each of which act on all atoms.
In summary, for each of \emph{compute\_[U,Y,dU,dE]} we launch a \emph{Kokkos::parallel\_for} in which each Kokkos thread (here mapping to a CUDA thread) performs work for a single atom.
For further details we refer the reader back to Sec.~\ref{sec:atom-loop-par} and to the reference code for \emph{compute\_U} in listing~\ref{lst:compute_u}.
In particular the Kokkos re-implementation of \emph{compute\_U} is unchanged relative to TestSNAP as the adjoint refactoring did not effect that kernel.

Because this stage is a prescriptive refactorization, success was evaluated by verifying correctness relative to the baseline. This is provided by construction in TestSNAP as discussed in Section~\ref{sec:SNAP}.
In contrast to the implementation before the adjoint refactorization demonstrated in \figurename~\ref{fig:atom-nbor-loop}, we already see a 15\% and 50\% improvement for the 2J8 and 2J14 problem sizes respectively.
We owe this benefit to the reduction in memory transfers, most importantly the elimination of the $\mathcal{O}(J^5)$ storage required for \emph{Zlist}.
Both problem sizes already show a speed-up relative to the baseline code.


\subsection {\textit{V2} - Atom and neighbor loop parallelization}
The next step is to expose additional parallelism over the neighbor dimension in the \emph{compute\_dU} and \emph{compute\_dE} routines.
We refer the reader back to Sec.~\ref{sec:neigh-loop-par} for the broad motivations.
In contrast to the TestSNAP case where we used an \emph{MDRangePolicy}, here we consider additional methods of parallelizing over the nested atom, neighbor loops.
One strategy to consider is assigning one Kokkos team, equivalently CUDA threadblock, to each atom and distributing the neighbors across the Kokkos threads, equivalently CUDA threads, within a block.
In our benchmark each atom has exactly 26 neighbors, implying one warp per threadblock (32 threads) would be sufficient.
Another strategy, the one we found most effective, is to collapse the neighbor and atom loops and launch a one-dimensional Kokkos \emph{parallel\_for}.
This allows Kokkos to manage scheduling work across threads and thread blocks.
We sketch this implementation in listing~\ref{lst:compute_du2}.

This strategy begged an opportunity to expose neighbor parallelism in the \emph{compute\_U} routine.
This requires the introduction of atomic additions.
As noted in Sec.~\ref{sec:SNAP}, for each atom the \emph{compute\_U} routine calculates the sum of expansion coefficients $\bU_j$ over each neighbor.
When we did not thread over neighbors this sum could be safely performed without atomics.
Our implementation featuring both atom, neighbor parallelism and atomic additions is given in listing~\ref{lst:compute_u2}.
\begin{minipage}{\linewidth}
\begin{lstlisting} [caption={Atom and neighbor loops collapsed in compute\_dU routine}, captionpos=b, label=lst:compute_du2]
void compute_dU
{
  const int nTotal = num_atoms*num_nbor;
  Kokkos::parallel_for(nTotal, KOKKOS_LAMBDA(int iter)
  {
    int natom = iter / num_nbor;
    int nbor = iter <@\%@> num_nbor;
    dUlist(natom,nbor,...) = ...
  });
}
\end{lstlisting}
\end{minipage}
\begin{minipage}{\linewidth}
\begin{lstlisting} [caption={2 kernels in compute\_U routine}, captionpos=b, label=lst:compute_u2]
void compute_U
{
  //Kernel 1
  Kokkos::parallel_for(nTotal, KOKKOS_LAMBDA(int iter)
  {
    int natom = iter / num_nbor;
    int nbor = iter <@\%@> num_nbor;
    for(int j = 0; j < idxz; ++j)
      Ulist(natom,nbor,j) = ... //Update Ulist
  });
  //Kernel 2
  Kokkos::parallel_for(nTotal, KOKKOS_LAMBDA(int iter)
  {
    int natom = iter / num_nbor;
    int nbor = iter <@\%@> num_nbor;
    for(int j = 0; j < idxz; ++j)
      Kokkos::atomic_add(Ulisttot(natom,j), Ulist(natom,nbor,j));
  });
}
\end{lstlisting}
\end{minipage}

In contrast to \textit{V1}, the evaluation of merit is non-trivial because
the atomic additions we introduced do have a performance penalty relative to non-atomic additions.
Nonetheless, profiler-driven optimization indicates this penalty is well amortized by the introduction of additional parallelism.
As documented in Figs.~\ref{fig:TestSNAP-progress-2J8} and~\ref{fig:TestSNAP-progress-2J14} we find a net $\sim$2$\times$ performance improvement from a verified correct introduction of neighbor parallelism for both the problem sizes.


\subsection {\textit{V3} - Data layout optimizations}
The previous two sections document exposing additional parallelism in each routine.
Exposing additional parallelism requires a careful reconsideration of data layouts to preserve sequential memory accesses across threads.
For GPUs, the data dimension over which the parallelization happens should be stored contiguously in memory. 
Our TestSNAP implementation originally used a CPU-friendly row-major data layout where the flattened $j$ indices were fastest while the parallelization happened over the atom dimension.
The strategy of threading over atoms lends itself to an atom-fastest data layout on GPUs.
Changing to a column-major data layout, and profiling each kernel to verify if the intended change resulted in memory coalescing, showed that this transformation only improved coalescing in \emph{compute\_Y}. The reason for this will be addressed in \textit{V4}.
For now, we note that this data layout modification in \emph{compute\_Y} gave us a 1.6$\times$ speedup for 2J8 and 1.3$\times$ speedup for the 2J14 problem sizes.
%

\subsection {\textit{V4} - Atom loop as the fastest moving index}
As mentioned earlier in \emph{compute\_[U,dU,dE]} there are still non-coalesced accesses because our convention for unflattening the atom, neighbor loop was neighbor-fastest.
The solution to this problem is trivially changing to atom-fastest.
Comparing the atom and neighor index calculation in listing~\ref{lst:compute_du2} and listing~\ref{lst:compute_du3}, shows us how the atom loop can be made the fastest index.
%
\begin{minipage}{\linewidth}
\begin{lstlisting} [caption={Atom loop as the fastest moving index}, captionpos=b, label=lst:compute_du3]
    int nbor = iter / num_atoms;
    int natom = iter <@\%@> num_atoms;
\end{lstlisting}
\end{minipage}
This loop reversal gave us another 2$\times$ performance boost.
We verified that these optimizations were implemented correctly by, as with \textit{V3}, profiling and checking memory coalescing metrics.
The data layout and loop order optimizations documented above gave an aggregate $\sim$3.5$\times$ speedup for 2J8 and $\sim$4$\times$ speedup for 2J14 problem size.
This is shown in bars \textit{V3} and \textit{V4} of \figurename~\ref{fig:TestSNAP-progress-2J8} and \ref{fig:TestSNAP-progress-2J14}.

\subsection {\textit{V5} - Collapse bispectrum loop}
At this point, the \emph{compute\_Y} routine was the most compute intensive. To expose more parallelism in the routine we assigned a single Kokkos team to an atom and distributed the flattened loops over $j$,$j_1$,$j_2$ across the Kokkos threads within a team.
As this optimization is prescriptive, success is defined by an improvement in performance.
We benchmarked this optimization and verfied a $\sim$ 3$\times$ performance boost in the particular kernel and an 80\% overall improvement in the performance as can be observed in the \textit{V5} bars of the performance figures.

\subsection {\textit{V6} - Transpose \emph{Ulisttot}}
Coalesced memory access is preferred on GPUs if consecutive threads access consecutive memory locations in a data structure.
We were aware at the conclusion of \textit{V3} that in \emph{compute\_Y} consecutive threads access strided indexes in \emph{Ulisttot}. This led to row-major being the optimal access pattern for \emph{Ulisttot} in \emph{compute\_Y}.
However its access in \emph{compute\_U} should be column-major to optimize the atomic add calls shown in listing \ref{lst:compute_u2}.
The solution to this problem is to add a transpose of \emph{Ulisttot} between \emph{compute\_U} and \emph{compute\_Y}.
This is a negligible overhead, 0.2\%, compared to the benefit of coalescing the repeated reads on \emph{Ulisttot}, giving a 15\% and 20\% performance improvement for 2J8 and 2J14 problem sizes.
This optimization is reflected in column \emph{V6} of Figs.~\ref{fig:TestSNAP-progress-2J8} and~\ref{fig:TestSNAP-progress-2J14}.

\subsection {\textit{V7} - 128 bit load/store}

Targeted profiling of TestSNAP using Nsight Compute showed that the \emph{SNAcomplex} complex data type was generating two distinct 64 bit loads and stores instead of a single 128 bit transaction.
This led to non-coalesced reads and writes on GPUs.
To guarantee the alignment requirements of 128 bit instructions we marked the \emph{SNAcomplex} with the C++11 \emph{alignas(16)} specifier.
This optimization gave an additional 15\% performance improvement for 2J8 and 19\% improvement for 2J14.

While TestSNAP gave us an overall of $7.5\times$ speedup for 2J8 and $8.9\times$ for 2J14, when these optimizations were implemented in the LAMMPS production code \cite{LAMMPS_testsnap}, we achieved a 7.3$\times$ speedup for 2J8 and an 8.2$\times$ speedup for 2J14 relative to the baseline.
\section{Architecture Specific Optimizations } \label{sec:snap2}

To this point each of our optimizations were agnostic to running on the CPU or GPU up to differences hidden within implicit conventions in the Kokkos framework. We identified that further optimizations were possible on the GPU relative to the CPU because of the required \emph{arithmetic intensity} (AI) to properly saturate each device. It is well known that GPU devices require a higher AI, equivalently FLOPS/byte fetched from memory, to break out of a memory bandwidth bound regime. In contrast
to the previous section, these optimizations were implemented directly in LAMMPS without first
being implemented in the TestSNAP proxy. Numerical correctness of the optimizations was verified by comparing the thermodynamic output (e.g. energy and pressure) of the new version to that of the baseline version over several timesteps.

\subsection {Recursive Polynomial Calculations}

Via a straightforward analysis, the kernels \emph{Compute\_U} and \emph{compute\_dU} should be (close to) compute-bound due to the large numbers of floating-point operations per atom, neighbor pair. However, GPU profiling found these kernels were memory bandwidth bound as implemented. After analysis, we determined this is due to repeated loads and stores to \emph{Ulist} and \emph{dUlist}, respectively, represented for example in listing~\ref{lst:compute_u2}. This is less of an issue on the CPU again because of the lower required arithmetic intensity to saturate the host processor. On the GPU, eliminating these loads and stores to device memory is a high priority optimization.

We will use the structure of the hyperspherical harmonic calculation as a motivator for our optimizations. The hyperspherical harmonics are defined by a recursion relation
\begin{equation}
\bu_j = \mathcal{F}(\bu_{j-\frac{1}{2}}) \label{eq:hysphrecur}
\end{equation}
where $\mathcal{F}$ is a linear operator in which each of the $(2j+1)^2$ elements of $\bu_j$  is a linear combination of two adjacent elements of $\bu_{j-\frac{1}{2}}$. The key observation is that this bounds the size of the state required for one atom, neighbor pair for any given value of $j$ to $16 \times (2j)^2$ bytes, the 16 corresponding to a complex double, from the previous level. This motivates using low-latency GPU \emph{shared memory} to cache the state, exposed in Kokkos as \emph{scratch memory}.

Shared memory is a limited resource per GPU compute unit, which imposes a strict bound on the maximum thread \emph{occupancy}, limiting the throughput of atom, neighbor pairs. This can be addressed using more CUDA threads per pair, both expediting the calculation and relaxing the occupancy limitations. Extra parallelism is afforded by the structure of (\ref{eq:hysphrecur}): for fixed $j$ the $(2j+1)^2$ elements of $\bu_j$ can be computed independently and thus concurrently.

Furthermore, we can exploit the symmetry property of $\bu_j$ to reduce the scratch memory requirements by roughly an additional factor of two. For fixed $j$, $\bu_j$ is dependent, up to a small overcounting for convenience of indexing, on $\mbox{ceil}(j + \frac{1}{2})$ rows of $\bu_{j-\frac{1}{2}}$. This reduces the scratch memory overheads to $(2j+1)(\mbox{floor}(j)+1)$ elements, again times two for the double buffer. Of important note, this memory layout optimization also carries over to \emph{Ulist} and \emph{Uarraytot}, roughly halving memory overheads, and also reducing the number of atomic additions required into \emph{Uarraytot}.

As a second, more nuanced optimization, we split the \emph{Uarraytot} data structure into two data structures corresponding to the real and imaginary parts. This is informed by the lack of ``\emph{double2}'' atomics, where two consecutive double precision values are updated atomically. While this removes the benefit of 128-bit loads and stores, this does allow non-strided atomic additions, net boosting performance.


An efficient implementation of this pattern is accomplished by assigning one GPU \emph{warp} per atom, neighbor pair, exposed in Kokkos via a \emph{VectorRange} construct. Here we parallelize computing the elements of $\bu_j$ over the ``vector lanes.''
The \emph{compute\_U} kernel in isolation achieves a 5.2$\times$ and 4.9$\times$ speedup for the 2J8 and 2J14 problem sizes, respectively, given these optimizations.

This implementation has additional benefits. Compared with listing~\ref{lst:compute_u2}, this implementation fuses the parallel loops for computing \emph{Ulist} and accumulating into \emph{Uarraytot}. This eliminates the need to store any value to \emph{Ulist}, allowing us to remove one of two data structures of size $\mathcal{O}(J^3 N_{\mathrm{atom}}N_{\mathrm{nbor}})$. Furthermore, now that one warp is assigned to each atom/neighbor pair, coalesced memory accesses require a return to a row-major or \emph{right\_layout} for \emph{Uarraytot}, removing the need for a transpose kernel before \emph{compute\_Y}.

We note that the number of components in a given $\bu_j$ is generally not divisible by 32, the warp length, and as such some threads will go unused. While this is not ideal, this load imbalance is a reasonable penalty to pay relative to the overall speed-up.

These optimizations transfer well to the \emph{compute\_dU} kernel. A recursion relation for $\bdu_j$ can be written from the derivative of (\ref{eq:hysphrecur}). There are two novel new constraints on the kernel. First, because we have eliminated \emph{Ulist}, we need to recompute values. This is net beneficial due to the relative cost of compute versus loads from memory. Next, because we need to compute \emph{dUlist} for \emph{x, y, z}, we need four times as much shared memory. Relative to \emph{compute\_U} this is a prohibitive occupancy limiter. The solution is to split \emph{compute\_dU} into three separate kernels, one per direction. Despite the redundant work this leads to a net speed-up.
%

We can eliminate the standing need to write \emph{dUlist} to memory via kernel fusion with \emph{update\_forces}. This requires re-introducing a global memory read from \emph{Ylist}, however the latency of this read can be hidden behind the computation of \emph{dU}; further, because \emph{Ylist} is a smaller data structure the reads are well cached. This is further improved by noting there is a one-to-one product between components of \emph{dU} and \emph{Ylist} in the kernel, letting us carry the symmetry properites of \emph{dU} over to \emph{Ylist}, again halving memory overheads. Last, because \emph{dE} is a function of atom, neighbor pairs, it can be accumulated within the kernel via a low-overhead \emph{parallel\_reduce} construct. For discussion we will refer to this new kernel as \emph{compute\_fused\_dE}.

The aggregate benefit of shared memory optimizations, splitting the kernels per-direction, halving memory overheads, and fusing in the force computation is a 3.3$\times$ and 5.0$\times$ speedup for the 2J8 and 2J14 problem sizes, respectively, for the new \emph{compute\_fused\_dE} kernel in isolation.

\subsection {Data Layout Optimizations}

A second point of optimization is improving the data layout of the \emph{Ylist} data structure. As described in \textit{V5}, the current data layout on the GPU has the flattened index $j,j_1,j_2$ as the fast-moving index and atom number as the slow index. While this led to a large improvement in performance, it was not without flaw.

As noted in (\ref{eqn:y}), $\bY$ is a sum over elements of $\bZ^j_{j_1j_2}$, which itself is a Clebsch-Gordon contraction of elements of $\bU$ as seen in (\ref{eqn:z}). The Clebsch-Gordon contractions are over a variable number of elements as a function of $j,j_1,j_2$. The convention of giving each thread of \textit{compute\_U} a different Clebsch-Gordon contraction implies both imperfect read coalescing and a natural load imbalance between different sums.

The former issue is a well-understood consequence of using an ``array-of-structures'' (AoS) data layout for \textit{Uarraytot} and \textit{Ylist}, where the ``S'' corresponds to the quantum numbers, while the ``A'' is for atom number. This layout was beneficial for \textit{Uarraytot} because, in the optimized implementation of \textit{compute\_U}, coalesced reads and writes over quantum numbers were naturally achieved---the ``S'' incidentally ended up acting as an ``A'' for the purpose of the kernel. On the other hand, \textit{compute\_Y} sees no such benefit. Profiling indicates that, while \textit{compute\_Y} does indeed leave read coalescing lacking, it still achieves good \emph{L1} cache reuse, which saves performance.

This leaves the problem of load imbalance between different sums. The solution to this problem is changing the data layout of \textit{Uarraytot} and \textit{Ylist} to an ``array-of-structures-of-arrays'' (AoSoA) data layout, and changing the threading pattern appropriately. For the new data layout, the inner-most (fastest index) ``A'' is fixed at 32---the length of a CUDA warp. This guarantees perfect read and write coalescing. The intermediate index ``S'' is the same as before---all quantum numbers. Last, the outer-most ``A'' is $\mbox{ceil}(N_{atom} / 32)$. A natural threading pattern follows, most easily achieved by a \textit{MDRangePolicy} as noted in listing~\ref{lst:compute-du-collapse}, except now it is a rank-3 policy mapping hierarchically to the data layout above.
Last, as \textit{compute\_Y} also requires atomic additions, we follow the pattern noted for  \emph{compute\_U} of spliting the AoSoA data structure into real and imaginary parts.
%

With this formulation, all reads and writes are perfectly coalesced, and there is perfect load balancing within a warp. 
For the \textit{compute\_Y} kernel in isolation, this optimization leads to a 1.4x speed-up for both the 2J8 and 2J14 problem sizes, owed to the elimination of load imbalance within a warp.

This reformulation to an AoSoA data layout does require the re-introduction of ``transpose'' kernels, as existed in optimization \textit{V6}. As was observed there, these additional transpose kernels are a negligible overhead relative to the improvement offered by the optimization of \textit{compute\_Y}.

\subsection {Overall Improvements}

\figurename~\ref{fig:snap-final} shows the final performance comparison of the newer implementation in LAMMPS \cite{LAMMPS_evan_update}  over the baseline after all the optimizations discussed in this paper. For the 2J8 benchmark, we see a 19.6$\times$ overall speedup over the baseline, whereas in the 2J14 case we observe a 21.7$\times$ speedup. Memory use was also greatly reduced: the 2J8 benchmark now uses only 0.1 GB of GPU memory, while the 2J14 benchmark uses 0.9 GB.
\begin{figure}
\centering
\includegraphics[width=1.0\columnwidth]{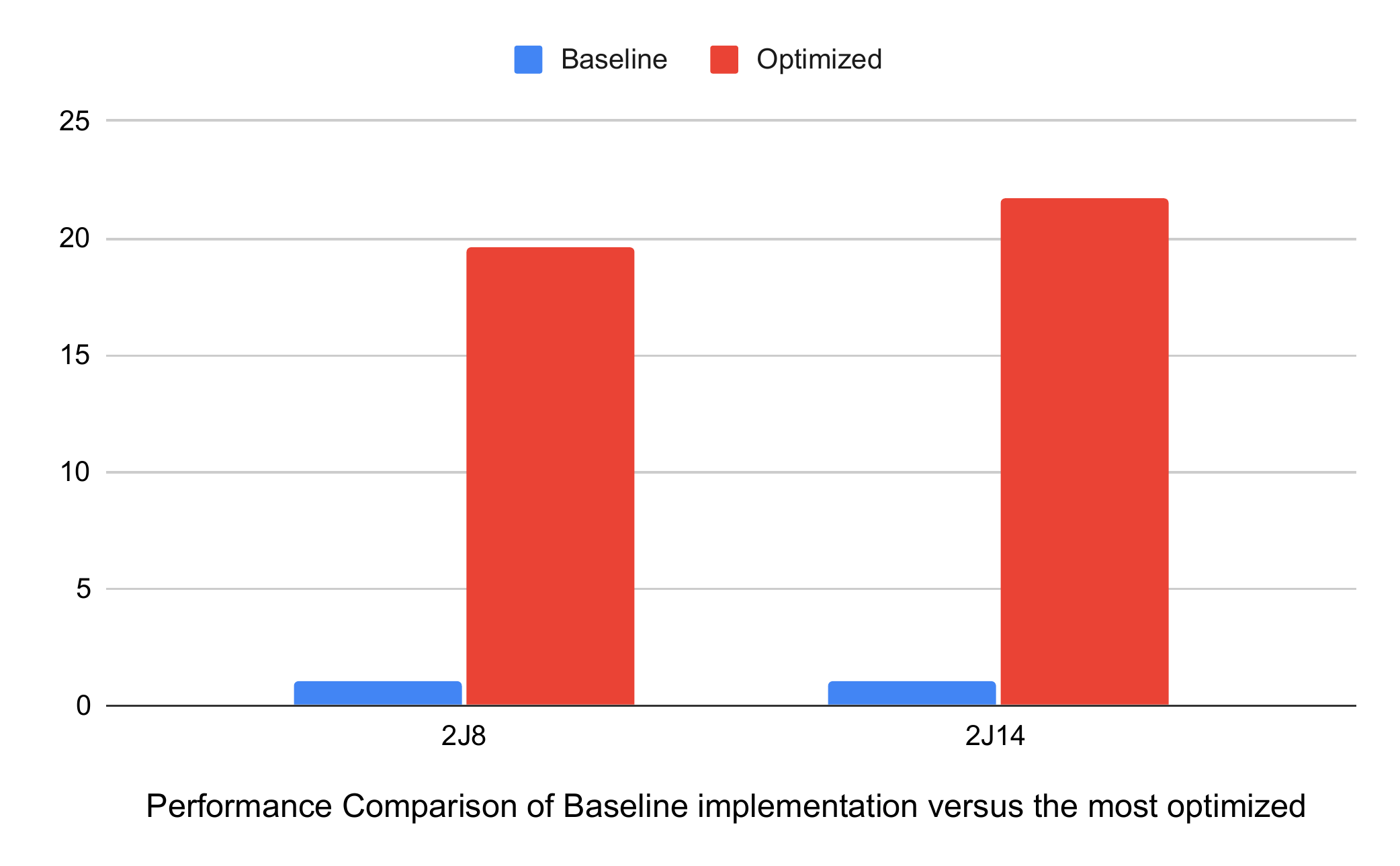}
\caption{Performance boost of new implementation compared to the older GPU implementation.}
\label{fig:snap-final}
\end{figure}

The fine-tuned optimizations for the GPU in this section lead to a performance regression on the CPU. For this reason, in our LAMMPS Kokkos implementation we used divergent code paths, where the CPU path largely uses only the optimizations up through section \ref{sec:TestSNAP2}.

We note that some aspects of the fine-tuned GPU optimizations do carry over to the CPU codepath. One is exploiting the symmetries of $\bu_j$, as well as the implicit extensions to $\bU_j$, $\frac{\partial \bU_j}{\partial {\bf r}_k}$, and $\bY$ via the one-to-one contraction with the derivative of $\bU_j$. This can still be used to half memory overheads across these data structures, which is important given the $\mathcal{O}(N_{atom} N_{nbor} J^3)$ memory scaling of \emph{Ulist} and \emph{dUlist} in particular. We note that backporting this optimization to the CPU (non-Kokkos) codepath has not yet occurred, but will be done in the future.

The second optimization of switching to an AoSoA data layout for \emph{Ylist} also has a generalization to the CPU, where instead of the inner-most ``A'' being of length 32, corresponding to the size of the warp, the inner-most ``A'' corresponds to the \emph{vector} index of a CPU SIMD data type. CPU SIMD optimizations to the SNAP potential will be explored in the future. 

\section{Conclusions}

In this paper we presented our efforts to improve the performance of the SNAP interatomic potential by using the NVIDIA V100-16GB GPU as the benchmark hardware.
We have quantified the benefits of a series of algorithmic and implementation strategies, leading up to an aggregate $\sim$22$\times$ speedup over a previous GPU implementation on the same hardware.
The fundamental adjoint refactorization of Sec.~\ref{sec:y-array} was essential for a viable implementation.
Key implementation optimizations included the counterintuitive approach of kernel fission instead of kernel fusion, loop reordering and data access optimizations, and architecture-specific optimizations such as shared memory caching of intermediate computations.
Each performance optimization in Sec.~\ref{sec:TestSNAP2} is inherently performance portable.
Additional shared memory optimization in Sec.~\ref{sec:snap2} are applicable across GPU architectures.
The majority of our algorithmic explorations were done in the standalone TestSNAP proxy-app which enabled the rapid conception and prototyping of many different optimizations.
We encourage all developers to develop proxy apps to shorten the design/implement/test/review cycle as was beneficial here.
Since then, all optimizations described here have been ported into the public release of the LAMMPS and are freely available.

\bibliographystyle{./IEEEtran}
\bibliography{IEEEabrv,references}



\end{document}